\documentclass[preprint, 5p, twocolumn]{elsarticle}

\usepackage{amssymb}

\usepackage{graphicx}
\usepackage{amsmath}
\usepackage{bbm}
\usepackage{xspace}
\usepackage{color}

\newcommand{\se}{Sec.\@\xspace}
\newcommand{\Se}{Section\@\xspace}

\newcommand{\ie}{i.\thinspace{}e.\@\xspace}

\newcommand{\ve}[1]{{\bf #1}}
\newcommand{\mat}[1]{\mathsf{#1}}
\newcommand{\nag}{{\phantom{\dag}}}

\newcommand{\eq}[1]{Eq.\thinspace{}(\ref{#1})}

\newcommand{\fig}[1]{Fig.\thinspace{}\ref{#1}}

\newcommand{\fc}[1]{({#1})}

\def\bra#1{\mathinner{\langle{#1}|}}
\def\ket#1{\mathinner{|{#1}\rangle}}
\def\braket#1{\mathinner{\langle{#1}\rangle}}

\journal{Computer Physics Communications}

\begin{document}

\begin{frontmatter}



\title{Polaritonic properties of the Jaynes-Cummings lattice model in two dimensions}


\author[graz]{Michael Knap\corref{cor1}}
\ead{michael.knap@tugraz.at}
\author[graz]{Enrico Arrigoni}
\author[graz]{Wolfgang von der Linden}
\address[graz]{Institute of Theoretical and Computational Physics, Graz University of Technology, 8010 Graz, Austria}
\cortext[cor1]{Corresponding author at: Institute of Theoretical and Computational Physics, Graz University of Technology, 8010 Graz, Austria. Tel.: +43 316 873 8189.}

\begin{abstract}
Light-matter systems allow to realize a strongly correlated phase where photons are present. In these systems strong correlations are achieved by optical nonlinearities, which appear due to the coupling of photons to atomic-like structures. This leads to intriguing effects, such as the quantum phase transition from the Mott to the superfluid phase. Here, we address the two-dimensional Jaynes-Cummings lattice model. We evaluate the boundary of the quantum phase transition and study polaritonic properties. In order to be able to characterize polaritons, we investigate the spectral properties of both photons as well as two-level excitations. Based on this information we introduce polariton quasiparticles as appropriate wavevector, band index, and filling dependent superpositions of photons and two-level excitations.  Finally, we analyze the contributions of the individual constituents to the polariton quasiparticles. 
\end{abstract}


\end{frontmatter}


\section{Introduction}
\label{sec:intro}
In the last few years proposals for new experimental realizations of
strongly-correlated many body systems emerged. Among them are
  ultracold gases of atoms trapped in optical lattices
  \cite{jaksch_cold_1998,greiner_quantum_2002,bloch_many-body_2008}
  and light-matter systems \cite{hartmann_strongly_2006,
    greentree_quantum_2006, angelakis_photon-blockade-induced_2007,
    hartmann_quantum_2008,tomadin_many-body_2010}. The latter consist
  of light modes which are confined in coupled cavity arrays. Due to
  the finite overlap of their quantum mechanical wave functions,
  photons are able to tunnel between adjacent cavities. Free photons
  are non-interacting, however, 
an effective
 repulsive interaction 
can be achieved by coupling photons to atoms or atomic-like structures
leading to interesting physical phenomena, which are subject to strong
correlations. 
Different theoretical models have been proposed
to describe these effects.
 In one scheme the atomic like structures are modelled by two level-systems, which leads to an interaction of the Jaynes-Cummings (JC) type \cite{jaynes_comparison_1963, greentree_quantum_2006}. Another scheme is based on four-level systems \cite{hartmann_strongly_2006} and takes advantage of electromagnetically induced transparency \cite{boyd_photonics:_2006}. Here, we focus on the first mentioned type of interaction. The physics of a single cavity at site $i$ is described by the JC Hamiltonian \cite{jaynes_comparison_1963}
\begin{equation}
 \hat{H}^{JC}_i = \omega_c \, a_i^\dagger \, a_i^\nag + \epsilon \, \sigma_i^+ \, \sigma_i^-+g( a_i^\nag\,\sigma_i^+ + a_i^{\dagger}\,\sigma_i^- ) \;\mbox{,}
 \label{eq:jc}
\end{equation}
where $\omega_c$ is the resonance frequency of the cavity, \ie, the frequency of the confined photons, $\epsilon$ is the energy spacing of
the two-level system, and $g$ is the atom-field coupling constant. The coupling between the atom and the photons is achieved by dipole interactions. The operator $a_i^\dagger$  ($a_i^\nag$) creates (annihilates) a photon at cavity $i$ and $\sigma_i^+$  ($\sigma_i^-$) is the raising (lowering) operator of the two-level system. The JC Hamiltonian conserves the particle number $\hat{n}_{i} = a_i^\dagger \, a_i^\nag + \sigma_i^+ \, \sigma_i^- $, which is a result of the rotating wave approximation \cite{haroche_exploringquantum:_2006}. The full system of coupled cavities is modelled by the Jaynes-Cummings lattice (JCL) Hamiltonian
\begin{equation}
 \hat{H}^{JCL} = -t \sum_{\left\langle i,\,j \right\rangle} a_i^\dagger \, a_j^\nag + \sum_i \hat{H}_i^{JC} - \mu\,\hat{N}_p \;\mbox{,}
 \label{eq:jcl}
\end{equation}
\begin{figure}
        \centering
        \includegraphics[width=0.35\textwidth]{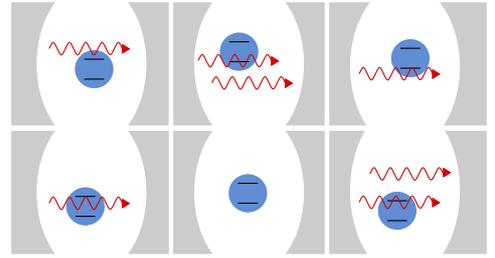}
        \caption{ Schematic illustration of the JCL model in two dimensions. Red wavy arrows indicate photons, which are confined in optical cavities and interact with two-level systems (blue bubbles). }
        \label{fig:jclfig}
\end{figure}
where the first term allows photons to tunnel between neighboring cavities $i$ and $j$. The restriction to nearest neighbors is indicated by the angle brackets $\langle \cdots \rangle$ around the summation indices. The hopping strength $t$ of the photons is given by the overlap integral of the photonic wave functions. The second term in \eq{eq:jcl} describes the JC physics of each cavity and the last term with the chemical potential $\mu$ controls the total particle number $\hat{N}_p = \sum_i \hat{n}_i$, which is a conserved quantity of the JCL Hamiltonian. An illustration of the two-dimensional JCL model is shown in \fig{fig:jclfig}.
Throughout this paper we use the dipole coupling $g$ as unit of energy. Under these considerations, the physics of the JCL model depends on three distinct parameters: the hopping strength $t$, the detuning $\Delta\equiv\omega_c-\epsilon$ and the modified chemical potential $\mu-\omega_c$.

The fundamental excitations present in the JCL system are termed polaritons. Polaritons are superpositions of both photons as well as two-level excitations. At zero hopping strength $t$ the JCL Hamiltonian reduces to the JC model with energies shifted by $-\mu \hat{n}_i$ and can thus be solved analytically \cite{jaynes_comparison_1963,hussin_ladder_2005,haroche_exploringquantum:_2006}. From this solution it follows that the energy which is necessary to add two polaritons to the cavity is larger than twice the energy which is necessary to add one polariton to the cavity, leading to a repulsive interaction between the particles. For small, non-zero hopping strength, the JCL system is in the so-called Mott phase, provided the polariton density is integer. In the case of light-matter systems the Mott phase can be regarded as a state of frozen light. For $t>0$ two energies are competing: The kinetic energy which is gained by the hopping process of the photons and the repulsive interaction between the particles. For $t$ larger than a critical hopping strength $t_c$, it is energetically favorable for the particles to delocalize on the whole lattice and to Bose condense in a state of zero momentum. In this parameter regime the JCL system is in the superfluid phase. The critical hopping strength $t_c$ defines the boundary of the quantum phase transition from the Mott to the superfluid phase.

Lately, there has been a great deal of research interest in understanding light-matter systems and in particular the JCL model. Most of the work has been devoted to study its ground state properties. The quantum phase transition from the Mott to the superfluid phase has been investigated by means of density matrix renormalization group \cite{rossini_mott-insulating_2007, rossini_photon_2008}, the variational cluster approach \cite{aichhorn_quantum_2008, knap_jcl_2010}, quantum Monte Carlo (QMC) \cite{zhao_insulator_2008}, and analytically by means of strong coupling perturbation theory \cite{schmidt_strong_2009, Schmidt_2010}. Results are also available on mean field level \cite{greentree_quantum_2006, koch_superfluid--mott-insulator_2009}; some signatures of the quantum phase transition have also been determined from exact diagonalization of small systems consisting of a few cavities \cite{angelakis_photon-blockade-induced_2007, makin_quantum_2008, irish_polaritonic_2008}. The spectral properties of the JCL model have been evaluated in Refs.~\cite{aichhorn_quantum_2008, pippan_excitation_2009, schmidt_strong_2009, knap_jcl_2010}.

In the present paper, we study in detail the polaritonic properties of the JCL model. This work extends our recent results published in Ref.~\cite{knap_jcl_2010} to two-dimensional systems. To be able to characterize polaritons, we first investigate the spectral properties of both particle species present in the JCL model, namely the photons as well as the two-level excitations. Based on this information we are able to introduce polariton quasiparticles as appropriate, wave vector, filling and band index dependent linear combinations of photons and two-level excitations. Furthermore we analyze the contributions of the individual constituents to the polariton quasiparticles. For completeness we present the boundary of the quantum phase transition from the Mott to the superfluid phase as well and compare our results to QMC results obtained in Ref.~\cite{zhao_insulator_2008}.

The remainder of this paper is organized as follows: in \se~\ref{sec:method} we present the numerical method we are using. \Se~\ref{sec:qpt} contains the results for the quantum phase transition and \se~\ref{sec:sp} provides a detailed investigation of the spectral properties of both particle species. Polaritonic properties are analyzed in \se~\ref{sec:pol}. Finally, we conclude and summarize our findings in \se~\ref{sec:conclusions}.

\section{Variational cluster approach}
\label{sec:method}
We employ the variational cluster approach \cite{potthoff_variational_2003, koller_variational_2006} (VCA) as a numerical tool to study the quantum phase transition, spectral properties and polariton quasiparticles of the two-dimensional JCL model. In particular, VCA provides the single-particle Green's function $G(\ve k,\,\omega)$ of the physical system $\hat{H}^{JCL}$ and is based on the self-energy functional approach \cite{potthoff_self-energy-functional_2003-1, potthoff_self-energy-functional_2003} and the cluster perturbation theory \cite{snchal_spectral_2000, snchal_cluster_2002}. VCA has been previously applied to light-matter systems in Refs.~\cite{aichhorn_quantum_2008, knap_jcl_2010, knap_tcl_2010}.
The main idea of VCA is that the 
self-energy $\mat \Sigma$ of the
physical system $\hat{H}^{JCL}$ is
approximated by the one of
a so-called reference system $\hat{H}^\prime$, which
shares its interaction part with $\hat{H}^{JCL}$ and is exactly 
solvable. Typically,  $\hat{H}^\prime $ is a
cluster decomposition of  $\hat{H}^{JCL}$.
The ``optimal'' reference system is determined by requiring that
 an appropriate functional $\Omega[\mat\Sigma]$ of the self-energy is stationary
(see \cite{potthoff_self-energy-functional_2003-1} for details).
$\Omega[\mat \Sigma]$ is the grand potential of the physical
system at the stationary point.
 Additionally, 
at the stationary point of $\Omega[\mat \Sigma]$, Dyson's equation is
recovered and thus the Green's function of the physical system can be
extracted there \cite{potthoff_self-energy-functional_2003-1}. 
 To evaluate $\Omega[\mat \Sigma]$ numerically, the self-energy is
 parametrized by the single-particle parameters $\mat x$ of the
 reference system $\hat{H}^\prime$. Due to this parametrization
 the functional $\Omega[\mat \Sigma]$ becomes a function $\Omega(\mat
 x)$, and is restricted to a smaller subspace of self-energies. 
Physical quantities are evaluated at the stationary point of $\Omega(\mat x)$ with respect to the single-particle parameters $\mat x$.  The accuracy of the results depends on the cluster size of the reference system, as correlations are taken into account exactly on the cluster level. Thus convergence on physical quantities is achieved by increasing the cluster size of the reference system. 

The reference system $\hat{H}^\prime$ is solved by exact diagonalization, which we carry out by means of the band Lanczos method \cite{freund_roland_band_2000, aichhorn_variational_2006}. To evaluate $\Omega(\mat x)$ and the Green's functions $G(\ve k,\,\omega)$ we apply the bosonic $\mat Q$-matrix formalism \cite{knap_spectral_2010}. In our previous work on the one-dimensional JCL model we adapted the VCA procedure such that it provides the single-particle Green's function of both particle species, which involved as a subtlety the mapping of the two-level systems onto hard-core bosons. To summarize, we are able to extract the Green's function of photons $G^{ph}(\ve k,\,\omega)$ and the Green's function of two-level excitations $G^{ex}(\ve k,\,\omega)$. From the Green's function $G^{x}(\ve k,\,\omega)$, where $x$ stands either for photons ($ph$) or for two-level excitations ($ex$), we obtain the single-particle spectral function
\begin{equation}
 A^x(\ve{k},\,\omega)\equiv-\frac{1}{\pi} \mbox{Im} \, G^x(\ve{k},\,\omega)
 \label{eq:spectralfunction}
\end{equation}
and the density of states
\begin{equation}
 N^x(\omega)\equiv \int A^x(\ve{k},\,\omega) \, d\ve{k}  = \frac{1}{N} \sum_{\ve{k}} A^x(\ve{k},\,\omega)\;\mbox{.}
 \label{eq:spe:dos}
\end{equation}

\section{Quantum phase transition}
\label{sec:qpt}
\begin{figure}
        \centering
        \includegraphics[width=0.35\textwidth]{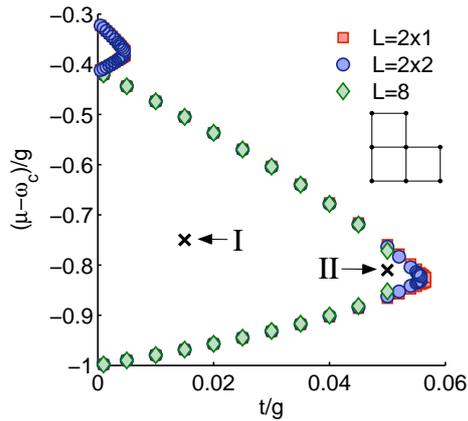}
        \caption{ Phase boundary of the two-dimensional JCL model for zero detuning $\Delta=0$ delimiting the first and the second Mott lobe from the superfluid phase. The phase boundary is evaluated for reference systems of size $L$. The inset shows the geometry of the 8 site cluster. The marks refer to parameters where spectral functions are evaluated. }
        \label{fig:pd}
\end{figure}

The two-dimensional JCL model exhibits a quantum phase transition from the Mott phase, where polaritons are localized within the cavities, to the superfluid phase, where polaritons delocalize on the whole lattice \cite{greentree_quantum_2006, aichhorn_quantum_2008, zhao_insulator_2008, schmidt_strong_2009}. We evaluate the phase boundary for zero detuning $\Delta=0$ by means of VCA from the minimal excitation energies of the system, see \fig{fig:pd}.  
\begin{figure}
        \centering
        \includegraphics[width=0.48\textwidth]{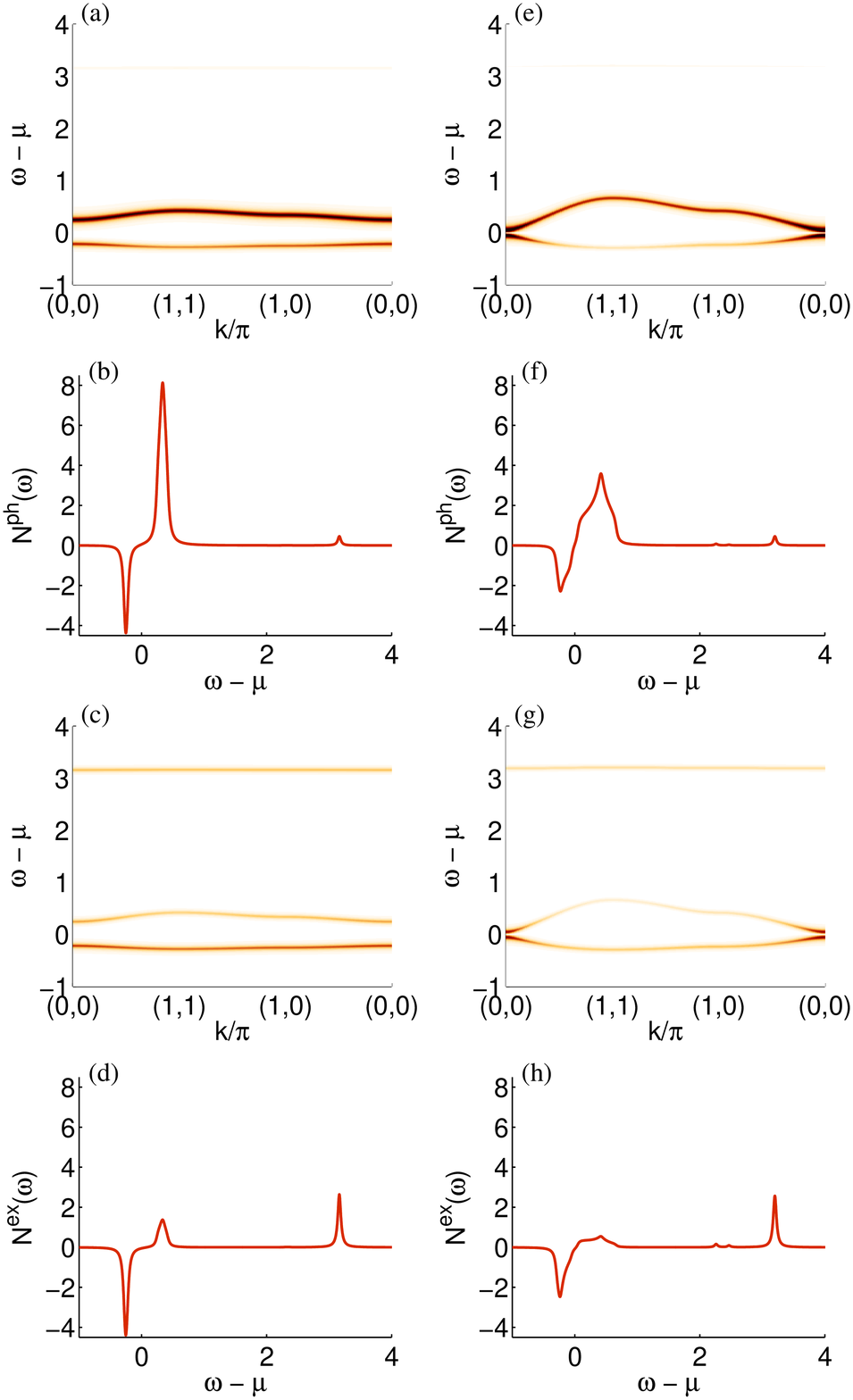}
        \caption{ Photon spectral function $A^{ph}(\ve{k},\,\omega)$, first row [\fc{a}, \fc{e}], and density of states $N^{ph}(\omega)$, second row [\fc{b}, \fc{f}]. Two-level excitation spectral function $A^{ex}(\ve{k},\,\omega)$, third row [\fc{c}, \fc{g}], and density of states $N^{ex}(\omega)$, fourth row [\fc{d}, \fc{h}]. The spectral functions are evaluated for the parameters \fc{a}--\fc{d} $t=0.015$, $\mu-\omega_c=-0.75$, $\Delta=0$, marked with $\mat x$ and with the Roman numeral I in \fig{fig:pd}, and \fc {e}--\fc{h} $t=0.05$, $\mu-\omega_c=-0.81$, $\Delta=0$, marked with $\mat x$ and with the Roman numeral II in \fig{fig:pd}. Both parameter sets belong to the first Mott lobe.}
        \label{fig:sf}
\end{figure}
Zero detuning means that the resonance frequency of the cavities is equal to the energy spacing of the two-level systems. We use the resonance frequency $\omega_c$ and the energy spacing $\epsilon$ of the two-level systems as variational parameters, \ie, $\mat x = \lbrace \omega_c,\,\epsilon \rbrace $, see  Ref.~\cite{knap_jcl_2010} for details concerning the choice of the variational parameters. Results converge quickly with increasing cluster size $L$. We evaluated the phase boundary for clusters consisting of $L=2\times1$, $L=2\times2$ and $L=8$ sites, where the latter are arranged according to the inset of \fig{fig:pd}. In the case of $L=8$ site clusters, the numerical determination of the stationary point of $\Omega(\mat x)$ turns out to be highly non-trivial. The reason for this behavior is subject to further research, which is out of the scope of this work. The width of the Mott lobes decreases with increasing particle density, which results from the fact that the repulsive interaction itself depends on the particle density. The critical hopping strength $t_c^n$, which specifies the tip of the $n$th Mott lobe, is $t_c^1 \approx 0.055$ for the first lobe and $t_c^2 \approx 0.0047$ for the second lobe. The VCA result for $t_c^1$ is in good agreement with the QMC data from Ref.~\cite{zhao_insulator_2008}, where the authors obtained $t_c^1\approx0.052$ for the critical hopping strength of the first Mott lobe.

\section{Spectral properties}
\label{sec:sp}
Spectral functions $A^x(\ve{k},\,\omega)$ and densities of states $N^{x}(\omega)$ evaluated for both particle species are shown in \fig{fig:sf}.
The spectral properties have been evaluated for the identical variational parameter set $\mat x = \lbrace \omega_c,\,\epsilon \rbrace$ as in the case of the quantum phase transition, for reference systems of size $L = 2 \times 2$ and an artificial broadening $0^+ = 0.03$. 

Spectral functions evaluated in the first Mott lobe consist of three bands. This can be understood from the zero hopping limit, where it is sufficient to consider a single cavity. The Hamiltonian then reduces to a block diagonal form, see for instance Refs.~\cite{jaynes_comparison_1963, haroche_exploringquantum:_2006} for a full solution of the single cavity problem. Each block corresponds to a specific polariton number, \ie, the states contributing to one block are $\ket{n,\,\downarrow}$ and $\ket{n-1,\,\uparrow}$, where the first quantum number denotes the number of photons and the second one states whether the two-level system is excited ($\uparrow$) or in its ground state ($\downarrow$). Diagonalizing the block yields the dressed states $\ket{n,\,\pm}$, where $\ket{n,\,-}$ is the ground state of the block and $\ket{n,\,+}$ its excited state. From these considerations it can be deduced that the single-particle excitations present in the spectral functions result from excitations from the ground state $\ket{n,\,-}$ to $\ket{n+1,\,-}$ or to $\ket{n+1,\,+}$. This yields two distinct particle bands, which we denote as $\omega_p^-$ and $\omega_p^+$, respectively. The same holds for the hole excitations, which corresponds to excitations from $\ket{n,\,-}$ to $\ket{n-1,\,-}$ or $\ket{n-1,\,+}$. We denote these bands as $\omega_h^-$ and $\omega_h^+$. However, for spectral properties evaluated in the first Mott lobe only one hole band exists, as the state $\ket{0,\,+}$ does not exist. This follows from the fact that in the zero hopping limit the block of zero particles consists only of one state $\ket{0,\,\downarrow}$. 

The location of the bands is identical for both the photons and the two-level excitations, as they are connected by the dipole coupling. However, the weights differ significantly. 
In particular, the particle band $\omega_p^-$ is much more pronounced in the photon spectral properties than in the two-level excitation spectral properties, whereas the opposite is true for the band $\omega_p^+$, which is almost not visible in $A^{ph}(\ve{k},\,\omega)$ but carries significant weight in $A^{ex}(\ve{k},\,\omega)$. The intensity of the hole band $\omega_h^-$ is similar in both cases.

\section{Polaritonic properties}
\label{sec:pol}

In this section, we investigate the polaritonic properties of the two-dimensional JCL model. To this end we follow Ref.~\cite{knap_jcl_2010}, and introduce polariton quasiparticle and quasihole creation operators as appropriate superpositions of photons and two-level excitations
\begin{align*}
 p_{\alpha,\ve{k}}^\dagger &= \beta_p^\alpha(\ve{k})\,a_\ve{k}^\dagger + \gamma_p^\alpha(\ve{k})\,\sigma^+_\ve{k} \;\mbox{,} \\
 h_{\alpha,\ve{k}}^\dagger &= \beta_h^\alpha(\ve{k})\,a_\ve{k} + \gamma_h^\alpha(\ve{k})\,\sigma^-_\ve{k} \;\mbox{.}
\end{align*}
The polariton quasiparticle states are defined as
\begin{align*}
 \ket{\tilde{\psi}_{p,\ve{k}}^\alpha} &= \frac{p_{\alpha,\ve{k}}^\dagger \ket{\psi_0} }{ \sqrt{\bra{\psi_0} p_{\alpha,\ve{k}} \, p_{\alpha,\ve{k}}^\dagger \ket{\psi_0} }} \;\mbox{and} \\
 \ket{\tilde{\psi}_{h,\ve{k}}^\alpha} &= \frac{h_{\alpha,\ve{k}}^\dagger \ket{\psi_0} }{ \sqrt{\bra{\psi_0} h_{\alpha,\ve{k}} \, h^\dagger_{\alpha,\ve{k}} \ket{\psi_0}}} \;\mbox{,}
\end{align*}
\begin{figure}
        \centering
        \includegraphics[width=0.48\textwidth]{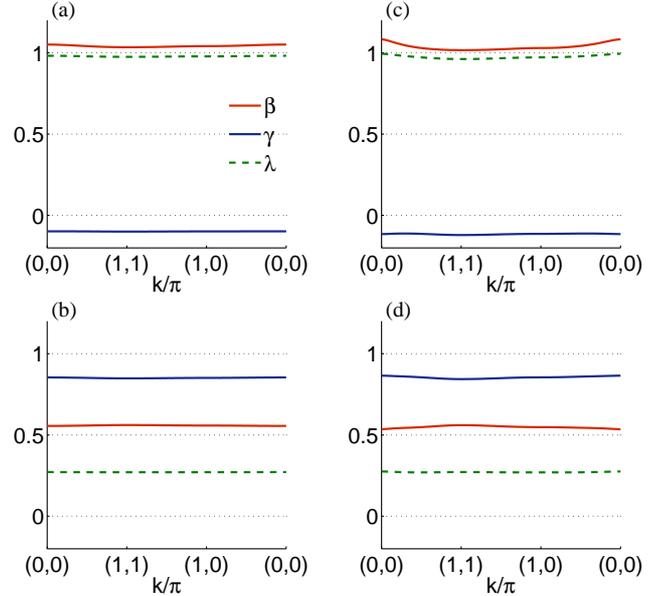}
        \caption{ Photon amplitudes $\beta$ and two-level excitation amplitudes $\gamma$ of the polariton quasiparticle creation operators \fc{a}, \fc{c} $p_{-,\ve k}^\dagger$ corresponding to the band $\omega_p^-$ and \fc{b}, \fc{d} $p_{+,\ve k}^\dagger$ corresponding to the band $\omega_p^+$. In addition, the overlap $\lambda$ is shown in all panels (dashed line). The left column, \fc{a}--\fc{b}, corresponds to the parameters $t=0.015$, $\mu-\omega_c=-0.75$, $\Delta=0$ [see mark I in \fig{fig:pd}] and the right column, \fc{c}--\fc{d}, to $t=0.05$, $\mu-\omega_c=-0.81$, $\Delta=0$ [see mark II in \fig{fig:pd}]. }
        \label{fig:qw}
\end{figure}
where polariton quasiparticle/quasihole operators are applied on the $N_p$-particle ground state $\ket{\psi_0}$. In the following arguments we neglect for simplicity the band index $\alpha$ and wave vector $\ve k$. The goal of this formulation is, to determine the coefficients of the linear combination $\beta$ and $\gamma$ such that the polariton quasiparticle states $\ket{\tilde{\psi}_{p/h}}$ describe best the true $N_p \pm 1$ particle system. Therefore the overlap of $\ket{\tilde{\psi}_{p/h}}$ with the true $(N_p\pm1)$-particle eigenstates has to be maximized with respect to the coefficients
\[\max_{\beta,\gamma} |\braket{\psi^{N\pm1}|\tilde\psi_{p/h}}|^2\;\mbox{.}\]
This condition yields a generalized eigenvalue problem \cite{knap_jcl_2010} of the form 
\[\mat A \, \ve z = \lambda \, \mat S \, \ve z\;\mbox{,}\]
where $\mat A$ is a $2\times 2$ matrix containing the spectral weight of photons and two level excitations contributing to the specific wave vector and band index, $\mat S$ is the overlap matrix and $\ve z$ determines the coefficients $\beta$ and $\gamma$ up to a constant, which is specified by the conservation of the total spectral weight. The eigenvalue $\lambda$ determines the quality of the polariton picture as it corresponds to the normalized overlap specified above. Thus $\lambda$ is bound by $[0,\,1]$. The polariton amplitudes $\beta$ and $\gamma$ of the linear combination are shown in \fig{fig:qw} for the particle bands $\omega_p^{\pm}$.
For the band $\omega_p^-$ the polaritons can be almost described by photonic excitations as $\beta$ dominates over $\gamma$. In addition the polariton picture is well fullfilled as $\lambda \approx 1$. For the band $\omega_p^+$ the situation is reversed as $\gamma$ dominates over $\beta$. In this case the polariton picture is solely modestly fullfilled as $\lambda \approx 0.3$.  Interestingly, for $\omega_p^-$ the amplitudes $\beta$ and $\gamma$ are of opposite sign, whereas for $\omega_p^+$ they are of the same sign. This behavior is already observed in the limit of a single JC cavity and is a result of the special structure of the eigenvectors \cite{knap_jcl_2010}. However, as compared to the single-cavity limit, in the coupled-cavity system the modulus of the amplitudes is significantly altered. Finally, in both cases $\omega_p^+$ and $\omega_p^-$, the polariton amplitudes $\beta$ and $\gamma$ depend slightly on the wave vector $\ve k$, which is also a pure lattice effect.

\section{Conclusions}
\label{sec:conclusions}

In the present paper, we investigated the polaritonic properties of the two-dimensional Jaynes-Cummings lattice model by means of the variational cluster approach. First, we evaluated the phase boundary delimiting the Mott from the superfluid phase and compared the critical hopping strength which determines the position of the Mott lobe tip with quantum Monte Carlo calculations. We found good agreement between the two approaches. In addition, we evaluated spectral functions and densities of states for both particle species present in the Jaynes-Cummings lattice model, namely photons as well as two-level excitations. Spectral functions generally consist of four bands, two particle bands and two hole bands. An exception are spectral functions evaluated for parameters of the first Mott lobe, which contain two particle bands but only one hole band. Furthermore, we investigated in detail the difference in the spectral weight of photons and two-level excitations. Based on the spectral properties of both particle species, we were able to introduce polariton quasiparticles as wave vector, band index and filling dependent superpositions of photons and two-level excitations. The amplitudes of the linear combinations determine the character of the polaritonic quasiparticles. Moreover, the amplitudes depend on the wave-vector, which is a pure lattice effect.

\section{Acknowledgments}
We made use of parts of the ALPS library \cite{albuquerque_alps_2007} for the implementation of lattice geometries and for parameter parsing.
We acknowledge financial support from the Austrian Science Fund (FWF) under the doctoral program ``Numerical Simulations in Technical Sciences'' Grant No. W1208-N18 (M.K.) and under Project No. P18551-N16 (E.A.).












\end{document}